\documentclass[aps,prb,twocolumn,superscriptaddress,amsmath,amssymb]{revtex4-2}

\usepackage{graphicx}
\usepackage{dcolumn}
\usepackage{bm}
\usepackage{subcaption}
\usepackage{epstopdf}
\usepackage{multirow}
\usepackage{footnote}
\begin{document}


\title{Structural and Optoelectronic Behaviour of Copper Doped $ Cs_{2}AgInCl_{6} $ Double Perovskite: A DFT Investigation}




\author{Isaac Busayo \surname{Ogunniranye}}
\email[Corresponding author: ]{ib.ogunniranye@ui.edu.ng}
\affiliation{Department of Physics, Faculty of Science, University of Ibadan, Ibadan, Nigeria.}

\author{Tersoo \surname{Atsue}}
\affiliation{Department of Physics, Faculty of Science, University of Ibadan, Ibadan, Nigeria.}
\affiliation{Department of Physics, Faculty of Physical Science,\\ Federal University Dutsin-Ma, Katsina, Nigeria.}

\author{Oluwole Emmanuel \surname{Oyewande}}
\email[Corresponding author: ]{oe.oyewande@ui.edu.ng}
\affiliation{Department of Physics, Faculty of Science, University of Ibadan, Ibadan, Nigeria.}

\date{\today}

\begin{abstract}
Recently, direct bandgap double perovskites are becoming more popular among photovoltaic research community owing to their potential to address issues of lead ($ Pb $) toxicity and structural instability inherent in lead halide (simple) perovskites. In this study, $ In-Ag $ based direct bandgap double perovskite, $ Cs_{2}AgInCl_{6} $ (CAIC), is treated with transition metal doping to improve the optoelectronic properties of the material. Investigations of structural and optoelectronic properties of $ Cu $-doped CAIC, $ Cs_{2}Ag_{(1-x)}Cu_{x}InCl_{6} $, are done using ab-initio calculations with density functional theory (DFT) and virtual crystal approximation (VCA). Our calculations show that with increasing $ Cu $ content, the optimized lattice parameter and direct bandgap of $ Cs_{2}Ag_{(1-x)}Cu_{x}InCl_{6} $ decrease following linear and quadratic functions respectively, while the bulk modulus increases following a quadratic function. The photo-absorption coefficient, optical conductivity and other optical parameters of interest are also computed, indicating enhanced absorption and conductivity for higher $ Cu $ contents. Based on our results, transition metal ($ Cu $) doping is a viable means of treating double perovskites - by tuning their optoelectronic properties suitable for an extensive range of photovoltaics, solar cells and optoelectronics.
\end{abstract}


\maketitle

\section{\label{sec:intro}INTRODUCTION}
Perovskite-based solar cells (PSCs) have recently been promoted as a renewable technology option for conventional solar cell technology capable of tackling global energy demands and climate change challenges owing to their economic and environmental viability \cite{Ibn-Mohammed2017}. Their emergence as one of the most promising emerging technologies has aroused the interest of the photovoltaic community, owing to their increasing power conversion efficiency (PCE) from $ 3.8 $\% in 2009 \cite{Kojima2009} to $ 25.2 $\% \cite{NREL2020}, materials availability, low production cost and ease of fabrication process \cite{Snaith2013,Liu2013,Burschka2013,Park2013,Di-Giacomo2014,Casaluci2015,Liu2016}. \\
One of the core elements governing PSC performance is the perovskite materials, typically $ CH_{3}NH_{3}PbI_{3} $ or MAPbI, serving both as light harvesters and charge carrier mediators \cite{Kojima2009,Liu2013,Ganose2017}. Over the years, numerous studies have shown the material (MAPbI) to possess appealing qualities needed for photovoltaic and optoelectronic applications, such as suitable bandgap ($ \sim 1.5 $ eV), good photoconductivity, considerable lifetime diffusion length ($ >100 $ $ \mu $m), high optical absorption coefficient, great bipolar transporting capability, defect tolerance ability, and low carrier effective masses with high mobility \cite{Fang2015,Baikie2013,Stranks2013,Hakamata2016,Even2013,Stranks2015,Watanabe2014,Sum2014,Edri2014,Walsh2015,Shao2014,Frost2014,Lindblad2014,Lin2015}. Despite having these exceptional properties, MAPbI still faces some fundamental issues; such as structural instability, toxicity associated with lead ($ Pb $), photocurrent hysteresis and scalability \cite{Shahbazi2016,Li2018b,Manser2016,Zhang2018b,Chatterjee2018,Niu2015}, which have hampered their large scale commercialization as viable PSCs. \\
Several attempts have been made to address the most fundamental issues of instability and toxicity inherent in halide perovskites; such as multi-cation substitution, hydrophobic moieties incorporation (e.g. hydrophobic polymer), surface passivation of perovskite absorber, carbon encapsulation, low dimensionality scheme/treatment and lead replacement with non-toxic elements \cite{Shahbazi2016,Wu2018,Tiong2018,Chen2018b,Zhang2018a,Wan2019,Dai2019,Johnson2019}. Reports have shown that substituting $ Pb $ with: (i) non-toxic group IVA elements ($ Sn $, $ Ge $) results to chemical instability and poor device performance owing to oxidation to their 4+ states \cite{Noel2014,Shao2018,Roknuzzaman2018,Kopacic2018}; (ii) isovalent elements ($ Bi $, $ Sb $) leads to reduced device efficiency \cite{Siddiqui2019}. This has thus created the need to develop novel classes of materials that are capable of tackling the instability and toxicity issues while still retaining the appealing properties of the $ Pb $-based halide perovskites (LHPs). \\
As of late, double perovskites (DPs) are beginning to gain popularity in the photovoltaic community owing to their ability to tackle issues of $ Pb $ toxicity and structural instability \cite{Roknuzzaman2019,Igbari2019}. DP is represented with the general $ A_{2}M^{'}M^{''}X_{6} $ stoichiometry, where $ A $ denotes large cation like $ Cs, Rb $; $ M^{'} $ and $ M^{''} $ represent monovalent and trivalent cations respectively ($ M^{'} = Ag^{+},Cu^{+}, Na^{+} ; M^{''} = In^{3+} , Bi^{3+} $) and $ X $ halide ($ X = Cl^{-}, I^{-}, Br^{-} $) \cite{Roknuzzaman2019,Igbari2019}. The materials properties of cation-ordered and vacancy-ordered DPs have been investigated experimentally and theoretically to determine their suitability for photovoltaic and optoelectronic applications. Research findings have shown that most DPs exhibit considerable thermal and mechanical stability compared to $ Pb $-based halide perovskites, but possess large bandgap with indirect nature, which have limited their usage in solar cell applications \cite{Roknuzzaman2019,Zhou2018,Deng2016,Wei2016,Zhao2018a}. \\
Given this, attentions are now drawn towards direct bandgap DPs. Direct bandgap DPs are in the forefront following the pioneering work by Volonakis and co-workers in 2017 where $ Cs_{2}AgInCl_{6} $ (CAIC) DP was proposed, synthesized and identified as a potential, environmentally-benign replacement for $ Pb $-based halide perovskites for photovoltaic and other optoelectronic applications \cite{Volonakis2017a}. CAIC is a direct-bandgap DP with high thermal and mechanical stability, which crystallizes in the face-centred cubic structure with space group $ Fm\bar{3}m $, and has an experimental lattice parameter of $ 10.469-10.481 $ \AA{} and bandgap of $ 2.5-3.3 $ eV \cite{Volonakis2017a,Zhou2017-1,Dahl2019}. These have made CAIC of huge research interest and a potential candidate for LHPs. However, pure bulk CAIC crystal or powder are characterized with low photoluminescence quantum yield (PLQY) and photo-absorption coefficient compared to CAIC nanocrystals (NCs) and these are as a result of parity-induced forbidden transition \cite{Zhou2017-1,Nandha2018,Liu2020-1}. \\
In a quest to tuning and optimizing the optoelectronic properties of DPs, experimental findings have identified doping engineering as a viable way of achieving this and thus, has the potential of enabling their widespread usability beyond photovoltaic applications \cite{Chen2019,Slavney2017}. Very recently, reports had shown the synthesis of doped-CAIC NC, treated with transition ($ Mn $) and post-transition ($ Bi $) metals, exhibiting high PLQYs, enhanced photo-absorption and other related optical properties when compared with pure CAIC either in powder or NC forms \cite{Nandha2018,Locardi2018,Luo2018,Liu2019}. \\
Numerical simulation has widespread application to a variety of problems \cite{Kolebaje2020,Oyewande2019a,Oyewande-Aisida2015,Yewande2009}. In particular, Monte Carlo simulation and mathematical modelling has been widely applied in the quest for more cost-effective fabrication of devices \cite{Akinpelu2019,Oyewande2019b,Kibbou2019,Yewande2006,Femi-Oyetoro2015,Akande2013a,Oyewande2018b,Oyewande2018,oyewande2012,Yewande2007,El2018,Yewande2005}, whereas properties of new materials have been studied with quantum mechanical calculations using density functional theory for decades \cite{Igbari2019,Johnson2019}. However, theoretical studies based on Density Functional Theory (DFT) on $ M $-cation doping in double perovskites are scarce. In a recent study, Jiao and associates used DFT scheme to investigate material properties of metal-alloying DP $ Cs_{2}AgM_{x}Br_{6} $ ($ M_{x} = Sb, In, Bi $), where indirect to direct bandgap transition was observed \cite{Jiao2019}. Transition metal doping can lead to enhancement of material properties, especially their optoelectronic properties. \\
Based on the scope of our literature search, $ Cu $-doping in CAIC is yet to be explored both theoretically and experimentally. Consequently, this paper seeks to investigate the effect of $ Cu $-doping on the structural and optoelectronic properties of CAIC ($ Cs_{2}Ag_{(1-x)}Cu_{x}InCl_{6} $) using the virtual crystal approximation (VCA) approach within the framework of DFT. VCA is a first-principles technique in modelling disordered solid solutions via pseudopotential averaging and effective in treating disordered systems \cite{Ramer2000b,Ramer2000a,Yu2007}. This work only focus on the bulk optoelectronic properties of perovskite materials. \\
The remaining part of paper is organized in the following order. In Section 2, the computational methods employed for the calculations are described. Section 3 is devoted to the presentation and discussion of our results. Finally, a brief summary of the work is given.

\section{\label{sec:methods}COMPUTATIONAL METHODS} 
In this work, the ab-initio calculations for $ Cs_{2}AgInCl_{6} $ (CAIC) and $ Cs_{2}Ag_{(1-x)}Cu_{x}InCl_{6} $ (CAIC:Cu) solid solutions were performed using the pseudopotential plane-wave technique based on density functional theory (DFT) as implemented in Quantum ESPRESSO (\textbf{QE}) software package \cite{Giannozzi2009,Giannozzi2017-1}. Within the framework of DFT, the structural and optoelectronic properties with the electronic exchange-correlation (XC) potential were calculated using Perdew-Berke-Ernzerhof (PBE) \cite{Perdew1996b} based on the generalized gradient approximation (GGA). The van der Waals functional (vdW-DF-OB86) \cite{Klimes2011} and the hybrid PBE0 functional \cite{Perdew1996a} were employed to treat the electronic exchange-correlation potential for the calculations of the lattice parameters and band structures respectively. \\
For the electron-ion interaction, the Optimized Norm-Conserving Vanderbilt (ONCV) pseudopotentials \cite{Hamann2013} were used for all calculations and construction of the virtual atoms ($ Ag_{(1-x)}-Cu_{x} $). The virtual crystal approximation (VCA) method \cite{Ramer2000b,Ramer2000a} was used to generate the pseudopotentials of the virtual atoms ($ Ag_{(1-x)}-Cu_{x} $), where the mixing ratio $ x $ was varied from 0 to 1 in the step of 0.1. The plane wave energy cut-off of $ 100 $ Ry and Monkhorst-Pack special \cite{Monkhorst1976} k-points of $ 6 \times 6 \times 6 $ were used for optimization calculations and the calculations of the electronic band structure and optical properties while a denser k-points of  $ 12 \times 12 \times 12 $ were used for  density of state (DOS) calculations. While the convergence threshold for self-consistent-field (SCF) iteration was set at $ 10^{-10} $ eV, the Broyden-Fletcher-Goldfarb-Shanno (BFGS) minimization method \cite{Fischer1992} was employed for the geometry optimization of the perovskites. The entire atomic positions were relaxed until the Helmann-Feynman forces on each atom become less than $ 20 $ meV/\AA{}. \\
To examine the optical properties of the perovskites, density functional perturbation theory (DFPT) \cite{Timrov2013,Timrov2015,Sharma2003} as implemented in \textbf{QE} was used to determine the complex frequency-dependent dielectric functions, $ \varepsilon (\omega) $:
\begin{equation}\label{eqn:eqn1}
\varepsilon(\omega) = \varepsilon_{1}(\omega) + i\varepsilon_{2}(\omega)
\end{equation}
From Eq. \ref{eqn:eqn1}, $ \omega $ denotes the photon frequency, $ \varepsilon_{1}(\omega) $ the real and $ \varepsilon_{2}(\omega) $ the imaginary parts of the dielectric function $ \varepsilon(\omega) $ respectively. To determine the light-harvesting capability of the perovskites, the absorption coefficients $ \alpha(\omega)  $ was calculated using Eq. \ref{eqn:eqn2}

\begin{equation}\label{eqn:eqn2}
\alpha(\omega) = \dfrac{\sqrt{2}\omega}{c}\sqrt{\left[ \mathbf{K} - \varepsilon_{1}(\omega) \right]}
\end{equation}
where
\begin{equation*}
\mathbf{K} = \sqrt{\varepsilon^{2}_{1}(\omega) + \varepsilon^{2}_{2}(\omega)}
\end{equation*}

Other optical parameters of interest are optical conductivity, refractive index, extinction coefficient and energy-loss function. The optical conductivity $ \sigma(\omega) $ and refractive index $ n(\omega) $ of the materials were computed using the relations in Eq. \ref{eqn:eqn3} and \ref{eqn:eqn3a} respectively.
\begin{equation}\label{eqn:eqn3}
\sigma(\omega) = \dfrac{\omega\varepsilon_{2}}{4\pi}
\end{equation}

\begin{equation}\label{eqn:eqn3a}
n(\omega) = \dfrac{1}{\sqrt{2}}\sqrt{\left[ \mathbf{K} + \varepsilon_{1}(\omega) \right]}
\end{equation}

In terms of the complex dielectric function in Eq. \ref{eqn:eqn1}, the extinction coefficient $ k(\omega) $ and the energy-loss function $ L(\omega) $ were determined by using Eq. \ref{eqn:eqn3b} and \ref{eqn:eqn3c} respectively.

\begin{equation}\label{eqn:eqn3b}
k(\omega) = \dfrac{1}{\sqrt{2}}\sqrt{\left[ \mathbf{K} - \varepsilon_{1}(\omega) \right]}
\end{equation}

\begin{equation}\label{eqn:eqn3c}
L(\omega) = \dfrac{\varepsilon_{2}(\omega)}{\mathbf{K}^{2}}
\end{equation}

\section{\label{sec:result_discussion}RESULTS AND DISCUSSION}
\subsection{Geometrical Stability}
Typical double perovskite structure is defined by the general $ A_{2}M^{'}M^{''}X_{6} $ stoichiometry. In this work, the metal cation in the crystallographic $ A $-site was taken to be $ Cs $, transition metal cation $ M^{'} = Ag,Cu,Ag_{(1-x)}Cu_{x} $ and post-transition metal cation $ M^{''}= In $ while the halide $ X = Cl $. To assert the crystallographic stability of CAIC:Cu solid solutions’ structure, we computed the perovskite formability parameters using Eq. \ref{eqn:eqn4} - \ref{eqn:eqn6}:

\begin{equation}\label{eqn:eqn4}
\mu = \dfrac{(r_{M'} + r_{M''})}{2r_{X}}
\end{equation}
\begin{equation}\label{eqn:eqn5}
t = \dfrac{(r_{A} + r_{X})}{\sqrt{2}\left[ \dfrac{(r_{M'} + r_{M''})}{2} + r_{X} \right]}
\end{equation}
\begin{equation}\label{eqn:eqn6}
\tau = \dfrac{2r_{X}}{(r_{M'} + r_{M''})} - n_{A}\left( n_{A} - \dfrac{2r_{A}/(r_{M'} + r_{M''})}{\ln(2r_{A}/(r_{M'} + r_{M''}))} \right) 
\end{equation}

where $ \mu $, $ t $ and $ \tau $ are octahedral , Goldschmidts tolerance and new tolerance factors respectively. $ r_{A} $, $ r_{M'} $, $ r_{M''} $, and $ r_{X} $ denote the Shannon radii \cite{Shannon1976} for the corresponding ions and $ n_{A} $ is the oxidation number of $ A $. For stable perovskite structures, the ideal range for $ \mu $, $ t $ and $ \tau $ are $ 0.44 \leq \mu \leq 0.9 $, $ 0.81 \leq t \leq 1.11 $ and $ \tau < 4.18 $ respectively \cite{Igbari2019,Meyer2018,Yu2019,Bartel2019}. The results in Table \ref{tab:tab1} indicate that the criteria for crystallographic stability for the halide double perovskites are satisfied. Thus, it can be inferred that CAIC and CAIC:Cu solid solutions will form stable three dimensional (3D) perovskite structures. This also indicates the feasibility of fabricating structurally stable CAIC:Cu solid solutions. 

 \begin{table}
 \caption{\label{tab:tab1} Perovskite formability factors for CAIC:Cu solid solutions.}
 \begin{ruledtabular}
 \begin{tabular}{|c|c|c|c|}
Material & $ \mu $ & $ t $ & $ \tau $ \\
\hline
$ Cs_{2}(Cu_{x}Ag_{1-x})InCl_{6} $ & 0.43 - 0.54 & 0.94 - 1.01 & -2.73 - (-2.40)
 \end{tabular}
 \end{ruledtabular}
 \caption*{where $ x $ is the mixing ratio in the step of ($ 0.1 \leq x \leq 1 $). }
 \end{table}
 
\subsection{Structural Properties}
\begin{figure}
 \centering
 \includegraphics[width=0.3\textwidth]{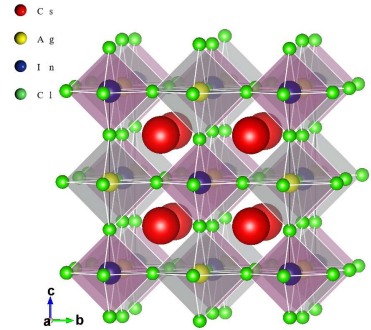}%
 \caption{\label{fig:fig1} : Polyhedral view of $ Cs_{2} AgInCl_{6} $ double perovskite (space group $ Fm\bar{3}m $).}
 \end{figure}
 
The host perovskite, CAIC, crystallizes in face-centred-cubic (fcc) phase with a space group of $ Fm\bar{3}m $ and its crystalline structure is illustrated in Fig. \ref{fig:fig1}. Within the framework of DFT, we used the van der Waals functional (vdW-DF-OB86) to accurately describe the lattice parameter ($ a $) and bulk modulus ($ B_{0} $) of the host perovskite by fitting the total energy-unit cell volume ($ E-V $) data into the Birch-Murnaghan equation of state (EOS) \cite{Birch1947}. The computed optimized lattice parameter for the host perovskite, CAIC, ($ 10.514 $ \AA{}) is found to be in good agreement with experimental values of $ 10.469 $ \AA{} \cite{Volonakis2017a} and $ 10.481 $ \AA{} \cite{Zhou2017-1}. The above procedure was repeated for $ Cs_{2}Ag_{(1-x)}Cu_{x}InCl_{6} $ while varying the $ Cu $-content $ x $ from 0 to 1 in the step of 0.1. To ascertain the reliability of the VCA method, the lattice parameter and electronic bandgap of $ Cs_{2}(Cu_{0.5}Ag_{0.5})InCl_{6} $ were computed using a $ 2 \times 2 \times 2 $ supercell. The results obtained show good agreement; $ 10.441 $ \AA{} ($ 10.450 $ \AA{}) and $ 0.408 $ eV ($ 0.336 $ eV) for VCA method (supercell alloying method) as presented in Table \ref{tab:tab2}.

\begin{table}
 \caption{\label{tab:tab2} Lattice parameter $ a $ and electronic bandgap $ E_{g} $ of $ Cs_{2}(Cu_{0.5}Ag_{0.5})InCl_{6} $ solid solution calculated using different alloying methods – virtual crystal approximation and supercell (SC).}
 \begin{ruledtabular}
 \begin{center}
 \begin{tabular}{cccc}
&& \multicolumn{2}{c}{\textbf{This work}} \\
     Material & Method & VCA & SC \\
      \hline
     \multicolumn{1}{c|}{\multirow{2}{*}{$ Cs_{2}(Cu_{0.5}Ag_{0.5})InCl_{6} $}} & $ a $(\AA{}) & $ 10.441 $ & $ 10.450 $ \\
     \multicolumn{1}{c|}{} & $ E_{g}(eV) $ & $ 0.408 $ & $ 0.336 $ \\
 \end{tabular}
 \end{center}
 \end{ruledtabular}
 \end{table}

\begin{figure}
 \centering
 \includegraphics[width=0.4\textwidth]{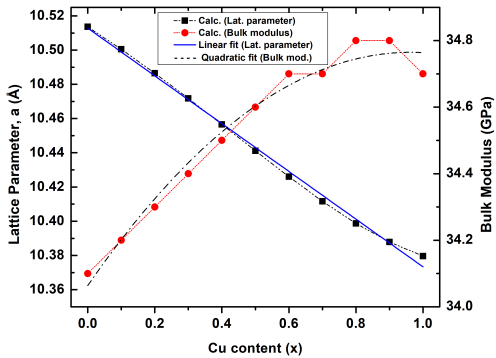}%
 \caption{\label{fig:fig2} : Calculated lattice parameters and bulk moduli as a function of $ Cu $-content $ x $ in CAIC:Cu solid solutions. Linear and quadratic fittings are presented.}
 \end{figure}

Fig. \ref{fig:fig2} shows the optimized lattice parameters and bulk moduli of CAIC:Cu solid solutions. It indicates an inverse relationship between the lattice parameter, $ a $ and $ Cu $-content $ x $. With an increase in $ Cu $-content $ x $, the lattice parameter decreases linearly with a function of $ a(x)=10.5128 - 0.1395x $, which satisfies the Vegard’s law for lattice parameters. Conversely, the bulk modulus ($ B_{0} $) of $ Cu $-doped CAIC tends to increase quadratically with a function of $ B_{0}(x) = 34.0636 + 1.4576 x - 0.7576x^{2} $. The implication of this is that the introduction of $ Cu $-dopant into CAIC does not only causes the crystal lattice to shrink but also reinforce the material stability. 

\subsection{Electronic Properties}
\begin{figure}
  \begin{subfigure}{.5\textwidth}
  \centering
  \caption{\label{fig:fig3}}
  \includegraphics[width=0.8\linewidth]{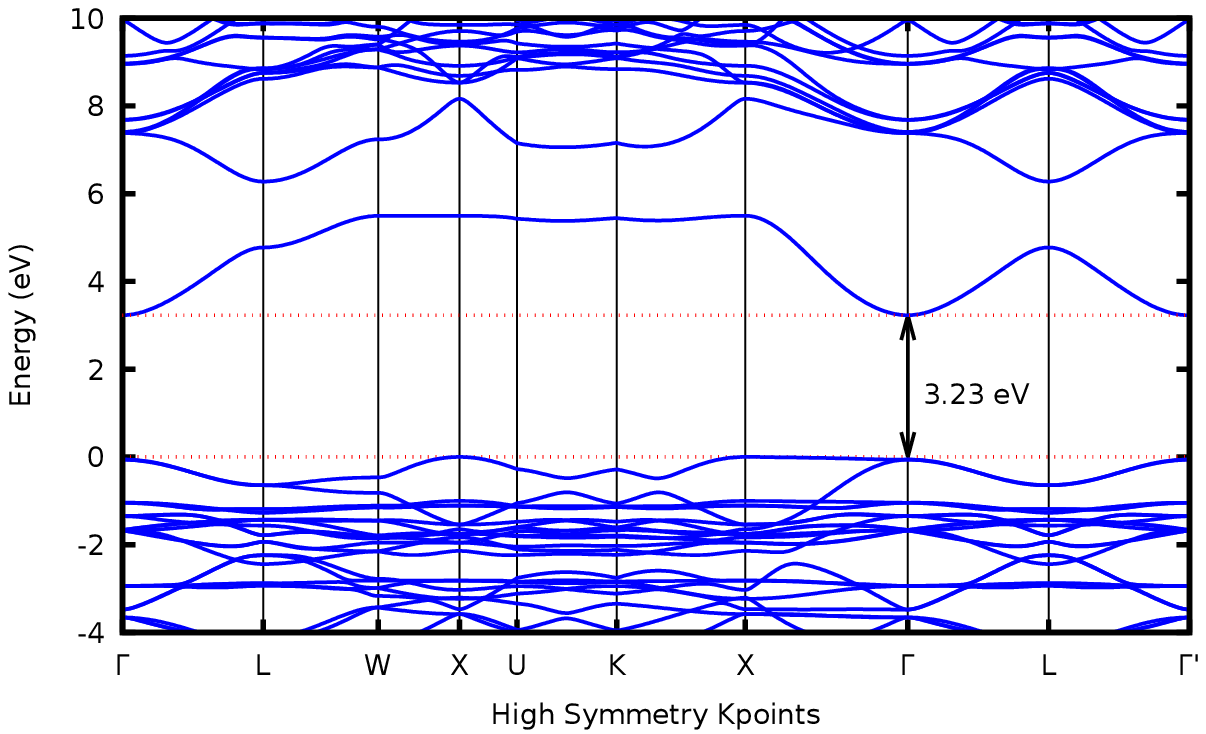} %
  \end{subfigure}%
. \begin{subfigure}{.5\textwidth}
  \centering
  \caption{\label{fig:fig4}}
  \includegraphics[width=0.8\linewidth]{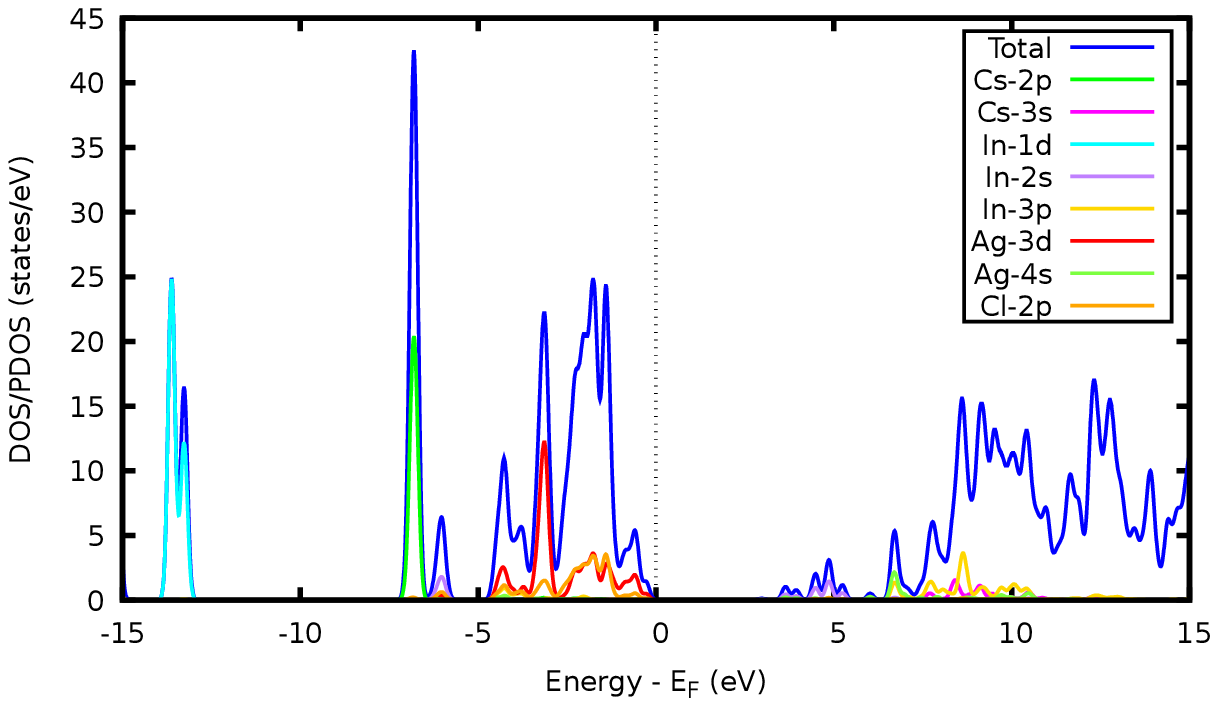} %
  \end{subfigure} %
  \caption{\label{fig:fig3_4} (a) The calculated PBE0 electronic band structure and (b) total and partial density of states (DOS/PDOS) of the host perovskite, $ Cs_{2}AgInCl_{6} $.}
\end{figure}

In this section, DFT based on the ab-initio calculations were used to examined the electronic structures of CAIC and CAIC:Cu solid solutions. From DFT calculations, we first employed GGA-PBE as the XC functional for the calculation of the bandgap and found that the calculated bandgap of CAIC ($ 0.95 $ eV) was about $ 70 $\% underestimated compared with the experimental values ($ 3.3 $ eV \cite{Volonakis2017a}, $ 3.23 $ eV \cite{Zhou2017-1}). Table \ref{tab:tab3} shows the comparative results of the calculated bandgap with other experimental and theoretical results. To circumvent this underestimation and improve the accuracy of the bandgap, we used the hybrid PBE0 as the XC functional and obtained a value ($ 3.23 $ eV) which is in a better agreement with experimental values (See Table \ref{tab:tab3}). At $ x=1 $, CAIC:Cu solid solution becomes $ Cs_{2}CuInCl_{6} $ (CCIC). \\
The nature of bandgap, as well as the positions of valence band minimum (VBM) and conduction band maximum (CBM), can be revealed via the electronic band structure. Fig. \ref{fig:fig3} shows the electronic band structure of CAIC (host perovskite) along some selected high symmetry points, which reflect that CAIC is a direct bandgap DP with both CBM and VBM located at the gamma ($ \Gamma $) point in the Brillouin zone. To ascertain the atomic orbital contributions towards the electronic states at CBM and VBM, the total and partial density of states (DOS) were calculated. Figure \ref{fig:fig4} shows the PBE0 total and partial density of states for CAIC where VBM is set at zero. From Figure \ref{fig:fig4}, the Ag-3d and Cl-2p states dominate the valence bands while the In-2s states exclusively dominate the conduction bands.

\begin{table}
 \caption{\label{tab:tab3} Calculated electronic bandgap $ E_{g} $ of CAIC and CCIC double perovskites using different exchange-correlation functionals compared with other experimental and theoretical results.}
 \begin{ruledtabular}
 \begin{center}
 \begin{tabular}{cccccc}
 && \multicolumn{2}{c}{\textbf{This work}} &&\\
      Material & & PBE & PBE0 & Previous work & Expt. \\
       \hline
      CAIC $ x=0 $ &\multicolumn{1}{|c|}{\multirow{2}{*}{$ E_{g}(eV) $}} & $ 0.95 $ & $ 3.23 $ & $ 2.9-3.3 $\cite{Volonakis2017a}, $ 3.33 $\cite{Zhou2017-1} & $ 3.3 $\cite{Volonakis2017a}, $ 3.23 $\cite{Zhou2017-1} \\
      CCIC $ x=1 $ & \multicolumn{1}{|c|}{} &  & $ 2.08 $ & $ 1.05\footnote[1]{HSE-SOC}-1.73\footnote[2]{PBE0-SOC; SOC - spin-orbit coupling.} $\cite{Pham2019} &  \\
  \end{tabular}
 \end{center}
 \end{ruledtabular}
 \end{table}

\begin{figure}
 \centering
 \includegraphics[width=0.4\textwidth]{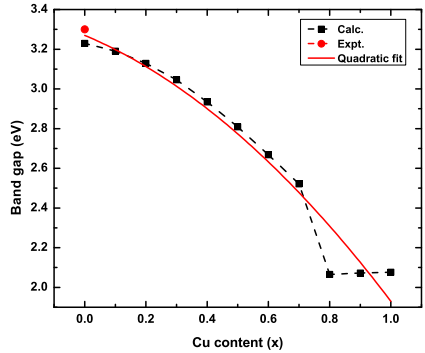}%
 \caption{\label{fig:fig5} Electronic bandgaps of CAIC:Cu solid solutions as a quadratic function of $ Cu $-content $ x $. Experimental bandgap value of CAIC is presented.}
 \end{figure}

Given the above, it is worth noting that the hybrid PBE0 functional can give the most reliable bandgap value for double perovskites (See Table \ref{tab:tab3}). With this assertion, the hybrid PBE0 functional was then used to calculate the electronic band structure of CAIC:Cu solid solutions. Fig. \ref{fig:fig5} shows the variation tendency in the bandgap as the $ Cu $-content ($ x $) increases. By interpolating the bandgaps to a polynomial function, the bandgaps were found to decrease quadratically with a function of $ E(x) = 3.2698 - 0.6463x - 0.6936x^{2} $, with increasing $ Cu $-content ($ x $). In addition to this, the direct bandgap nature of the host perovskite remains unchanged despite the addition of $ Cu $-dopants. Based on the Vegard’s law, this function can be reduced to:

\begin{equation}
E_{g}(x) = E_{g}(0) + \left[ E_{g}(1) - E_{g}(0) - b \right]x + bx^{2}  
\end{equation}

where $ E_{g}(0) $ and $ E_{g}(1) $ denote the bandgap of CAIC ($ x=0 $) and CCIC ($ x=1 $) respectively, while $ b $ represents the band gap bowing parameter. From our results, $ E_{g}(0) = 3.2698 $ eV, $ E_{g}(1) = 1.9299 $ eV and $ b = -0.6936 $ eV. Bandgap bowing parameter ($ b $) indicates the non-linearity of the bandgap to the composition, as well as the degree of fluctuation in the crystal field. The results imply enhancement in the light-absorbing capability of the materials owing to the reduction in the bandgap with increasing $ Cu $-content. Since the band gap bowing parameter ($ b=-0.6936 $) is very small, it indicates good miscibility between CAIC and CCIC, and low compositional disorder.

\begin{figure}
  \begin{subfigure}{.5\textwidth}
  \centering
    \caption{\label{fig:fig6}}
  \includegraphics[width=0.8\linewidth]{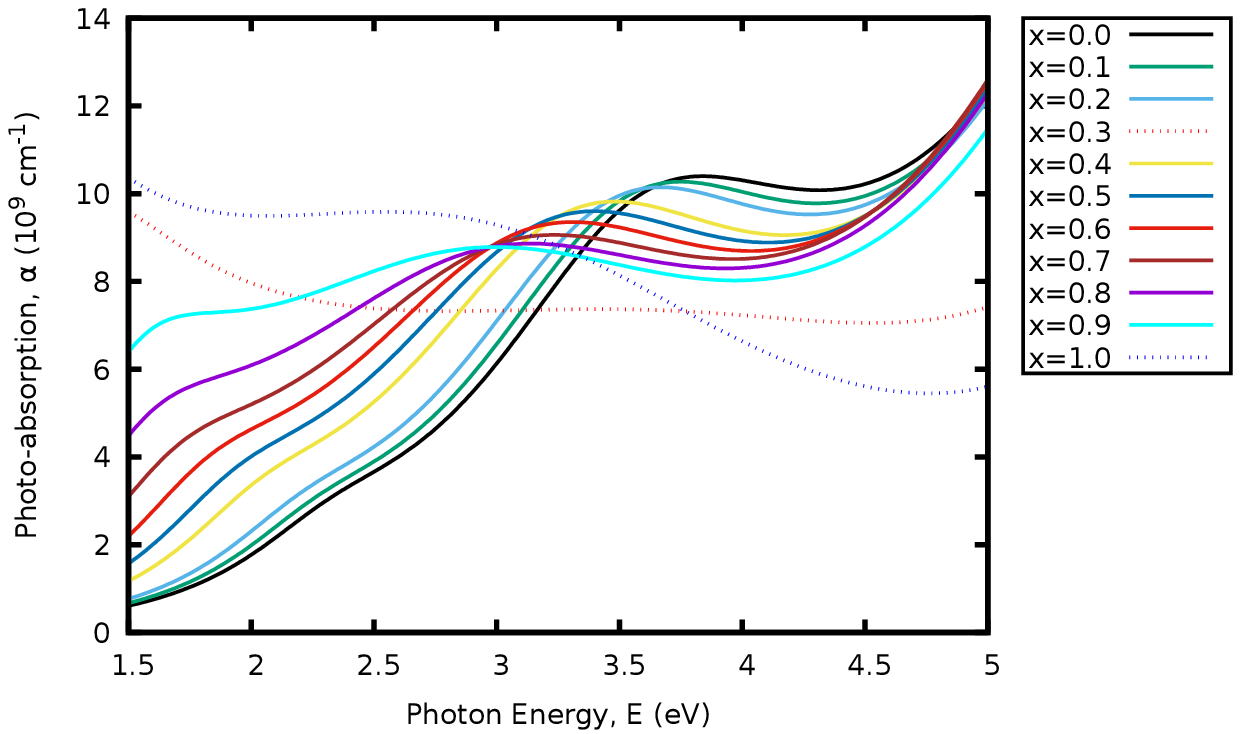} %
  \end{subfigure}%
. \begin{subfigure}{.5\textwidth}
  \centering
  \caption{\label{fig:fig7}}
  \includegraphics[width=0.8\linewidth]{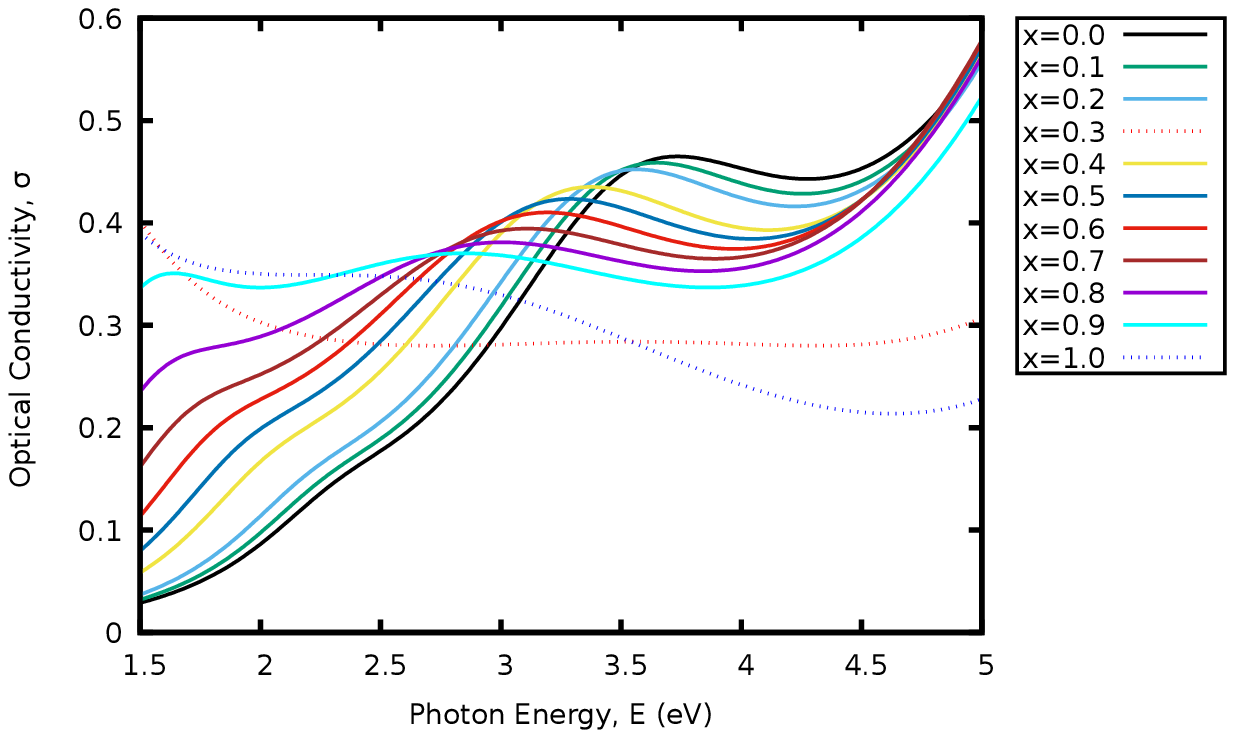} %
  \end{subfigure}%
  \caption{\label{fig:fig6_7} Optical properties of $ Cu $-doped CAIC double perovskites with the GGA-PBE functional. (a) Calculated photo-absorption coefficients. (b) Calculated optical conductivity.}
\end{figure}

\begin{figure}
 \centering
 \includegraphics[width=0.4\textwidth]{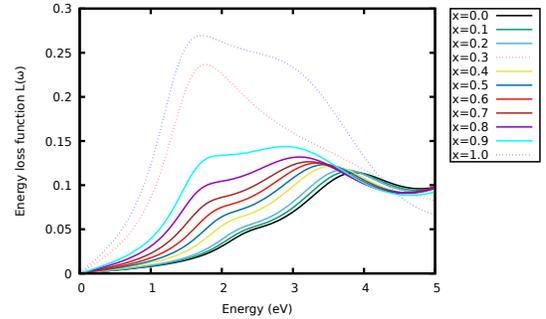}%
 \caption{\label{fig:fig8} Calculated energy-loss function $ L(\omega) $ for CAIC:Cu solid solutions.}
 \end{figure}

\subsection{Optical Properties}
In this section, we focus on the optical properties within the visible range ($ 1.5-3.1 $ eV), where the majority of solar energy lies. To quantify the effects of $ Cu $-doping on the optical properties, the photo-absorption coefficients and optical conductivity of CAIC:Cu solid solutions are calculated and presented in Fig. \ref{fig:fig6_7}. One of the key parameters that determine the power conversion efficiency of PSCs and other related optoelectronic devices is the photo-absorption coefficient. Photo-absorption coefficient gives one good insight into the light-harvesting capability of a material. It  can be estimated using Eq. \ref{eqn:eqn2} with the real and imaginary parts of the complex frequency-dependent dielectric functions. Fig. \ref{fig:fig6} shows the variation of the photo-absorption coefficients for CAIC and CAIC:Cu solid solutions as a function of photon energy. Within the visible range, the photo-absorption coefficients steadily increase with energy. The absorption coefficients of CAIC:Cu solid solutions are larger than that of CAIC. In comparison with CAIC, increased photo-absorption coefficients of CAIC:Cu are notably observed across the whole visible range. In particular, CCIC shows maximum absorption in the energy range of $ 1.5-2.7 $ eV. With these results, it can be inferred that the incorporation of $ Cu $-dopant can increase the photo-absorption coefficients of CAIC. Thus, CAIC:Cu solid solutions exhibit good absorption within the visible region. 
   
   \begin{figure}
     \begin{subfigure}{.5\textwidth}
     \centering
       \caption{\label{fig:fig9}}
     \includegraphics[width=0.8\linewidth]{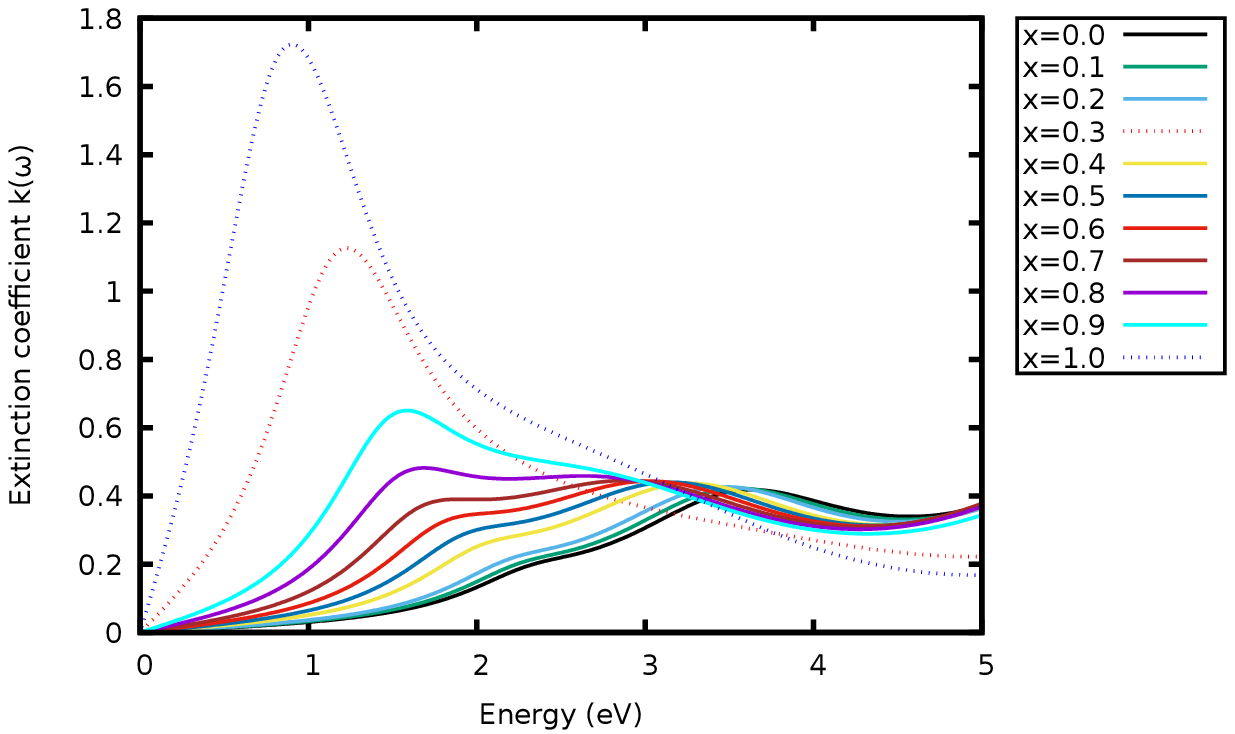} %
     \end{subfigure}%
   . \begin{subfigure}{.5\textwidth}
     \centering
     \caption{\label{fig:fig10}}
     \includegraphics[width=0.8\linewidth]{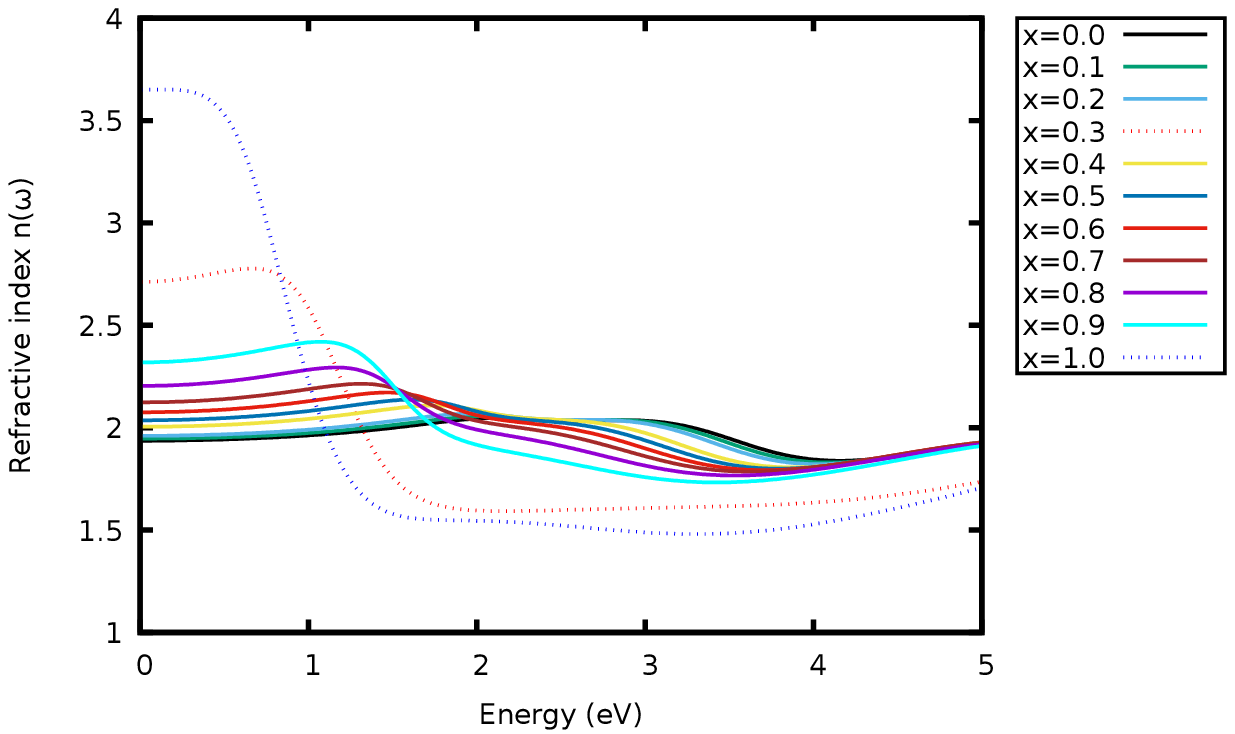} %
     \end{subfigure}%
     \caption{\label{fig:fig9_10} Variation in the calculated (a) Extinction coefficient $ k(\omega) $ and (b) Refractive index $ n(\omega) $ as a function of energy for CAIC:Cu solid solutions.}
   \end{figure}

Furthermore, the optical conductivity of the materials was also examined. Fig. \ref{fig:fig7} shows the variation of the calculated optical conductivity of the materials. A similar trend was observed as in Fig. \ref{fig:fig6}. Other optical parameters for the materials were computed and the results presented in Fig. \ref{fig:fig8} - \ref{fig:fig9_10}. These show the variation of the calculated energy-loss function, extinction coefficient and refractive index as a function of energy. From the calculated spectra in Fig. \ref{fig:fig8} and \ref{fig:fig10}, the extinction coefficient and energy-loss function display peak values of 1.13 at 1.26 eV, 1.73 at 0.92 eV, 0.25 at 1.79 eV and 0.28 at 1.79 eV. With these findings, increasing the concentration of $ Cu $-content in CAIC does not only enhance the light-harvesting ability of the material but also lead to enhancement of the device efficiency by extension. 

\section{\label{sec:conclusion}CONCLUSION}
In this work, the effect of $ Cu $-doping on the structural and optoelectronic properties of CAIC has been studied using first-principles DFT calculations and VCA approach. The ab-initio VCA method was used to model the solid solutions. The PBE0 functional was used for the band structure calculations after assessing the exchange-correlation functional of GGA-PBE. With increasing $ Cu $-content, the crystal lattice shrinks following a linear function $ a(x) = 10.5128 - 0.1395x $, bulk modulus increases with a quadratic function of $ B_{0}(x) = 34.0636 + 1.4576x - 0.7576x^{2} $, while the bandgap decreases quadratically with a function of $ E(x) = 3.2698 - 0.6463x - 0.6936x^{2} $. The photo-absorption coefficient, optical conductivity and other optical parameters of interest are calculated using the DFPT method. The spectra obtained show enhanced absorption and conductivity at higher $ Cu $-content. The variation tendencies, as a result of $ Cu $-doping, in the structural and optoelectronic properties of the materials under study have shown $ Cu $ to be an efficient dopant in treating double perovskites. This work presents $ Cs_{2}Ag_{(1-x)}Cu_{x}InCl_{6} $ (CAIC:Cu) solid solutions as potential candidates for photovoltaics and optoelectronics.

\begin{acknowledgments}
The authors wish to acknowledge The Postgraduate College of the University of Ibadan, Ibadan, Nigeria for the computational access to her High-Performance Computing (HPC) resources. All the computational tasks were performed with these resources.
\end{acknowledgments}

\bibliography{my-references}

\end{document}